\documentclass[a4paper,prl,twocolumn,showpacs]{revtex4}

\usepackage{amssymb}
\usepackage{amsfonts}
\usepackage{amsmath}
\usepackage{color}
\usepackage{graphics, graphicx}

\newcommand{\ff}[1]{{\boldsymbol #1}}

\begin{document}

\author{Andrej Schwabe}
\author{Daniel G\"utersloh}
\author{Michael Potthoff}

\affiliation{I. Institut f\"ur Theoretische Physik, Universit\"at Hamburg, Jungiusstra\ss{}e 9, 20355 Hamburg, Germany }

\pacs{75.75.-c,75.10.Jm,72.10.Fk,75.30.Et,85.75.-d} 


\title{Competition between Kondo screening and indirect magnetic exchange in a quantum box}

\begin{abstract}
Nanoscale systems of metal atoms antiferromagnetically exchange coupled to several magnetic impurities are shown to exhibit an unconventional re-entrant competition between Kondo screening and indirect magnetic exchange interaction.
Depending on the atomic positions of the magnetic moments, the total ground-state spin deviates from predictions of standard Ruderman-Kittel-Kasuya-Yosida perturbation theory.
The effect shows up on an energy scale larger than the level width induced by the coupling to the environment and is experimentally accessible by studying magnetic field dependencies.
\end{abstract}

\maketitle

\paragraph{Introduction.}

The competition between different mechanisms working at comparable energy scales is in many cases the origin of unconventional physical properties. 
The Kondo temperature $T_{\rm K}$ is the energy scale below which the local magnetic moment of a magnetic impurity  is screened by the conduction-electron spins of a metallic host \cite{Hew93}.
In multi-impurity systems, the Kondo effect competes with inter-impurity magnetic correlations caused by the Ruderman-Kittel-Kasuya-Yosida indirect magnetic exchange interaction \cite{RK54,Kas56,Yos57}.
As was already pointed out by Doniach in 1977 \cite{Don77}, their different scaling with the antiferromagnetic coupling $J$, namely  $T_{\rm K} \sim e^{-1/J}$ and $J_{\rm RKKY} \sim J^2$ (for weak $J$), gives rise to a point $J=J_{\rm D}$ where $T_{\rm K} = J_{\rm RKKY}$. 
This roughly marks a crossover or even a phase transition and represents an important key to understand the phase diagrams of dilute Kondo systems or Kondo lattices.
Here we ask: How does this competition between the Kondo screening and RKKY interaction change if the quantum system is made so small that its conduction-electron spectrum becomes discrete? 
How does the physics change due to the presence of a third energy scale, the level spacing $\Delta$ close to the Fermi energy?

For a {\em single} magnetic impurity, i.e.\ for the ``Kondo-box'' problem \cite{TKvD99}, there is already a competition, namely between $\Delta$ and $T_{\rm K}$ \cite{Sch01,Sch02,SA02,CB02,FRF03,HKM06,BBH10}:
If the level spacing $\Delta$ becomes comparable to the bulk $T_{\rm K}$, logarithmic Kondo correlations are cut, and the extension of the Kondo screening cloud is actually given by the system size.
There is eventually only a single conduction-electron state within the Kondo scale $T_{\rm K}$ around the Fermi energy which is available to form the ``Kondo'' singlet.
Here, we will argue that this feature results in an unconventional, spatially dependent competition with the RKKY interaction for the multi-impurity case.
This becomes relevant for magnetic nanostructure physics in an important parameter range and for the bottom-up construction of spintronics devices.

\begin{figure}[b]
\vspace{-4mm}
\centerline{\includegraphics[width=0.5\textwidth]{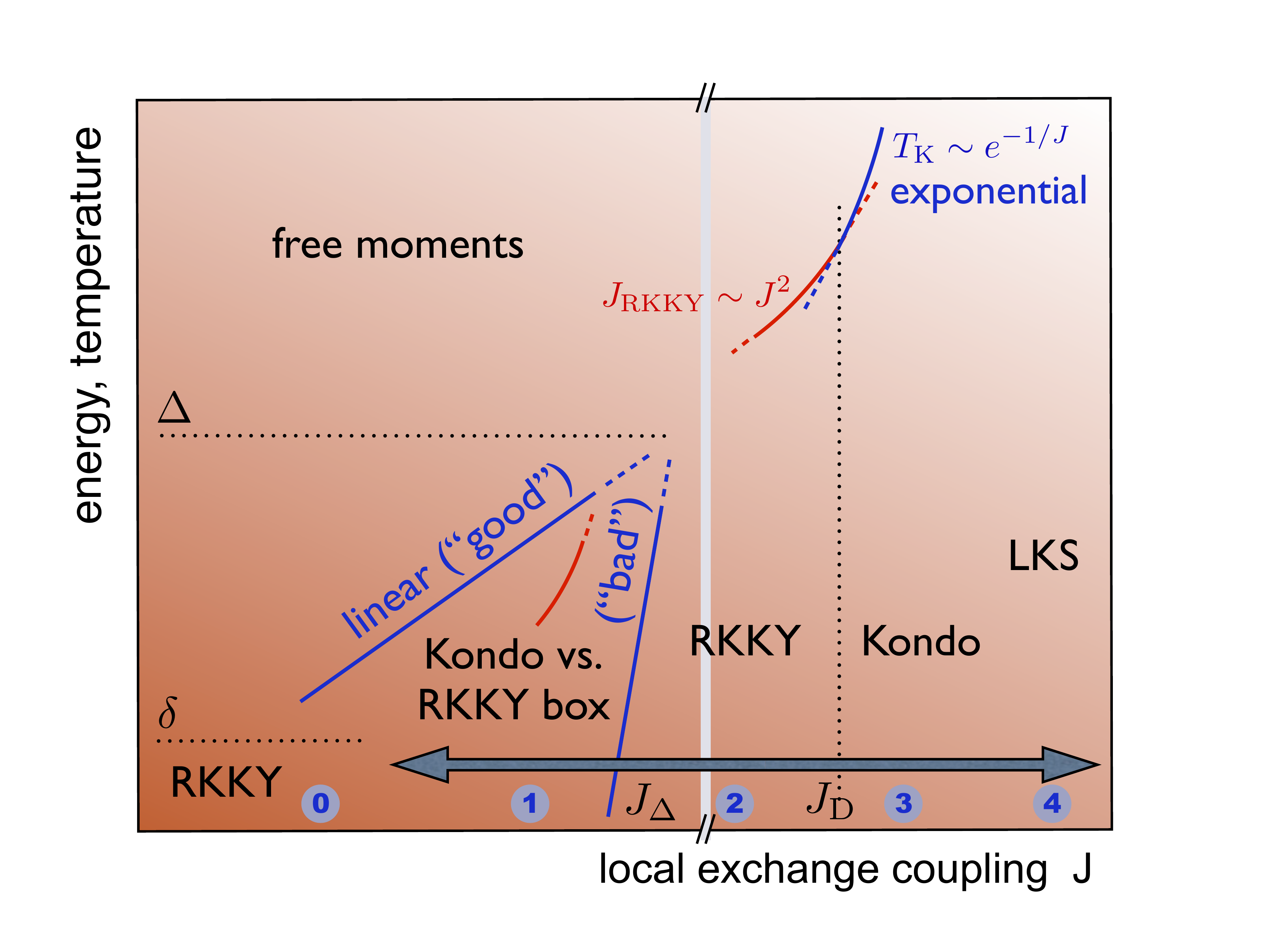}}
\vspace{-4mm}
\caption{(Color online) 
Competition between Kondo screening and RKKY interaction in a quantum box.
Local Kondo singlets (LKS) are formed for extremely strong $J$ (regime 4).
With decreasing $J$, the energy to break up a Kondo singlet becomes exponentially small (3).
Below $J_{\rm D}$, RKKY coupling is dominant (2).
Finite-size effects set in for $J<J_{\Delta}$ defined as the point where the bulk Kondo temperature equals the finite-size gap, $T_{\rm K}=\Delta$.
If the system is sufficiently large, we have $J_{\Delta} < J_{\rm D}$.
The singlet-formation energy is linear in $J$ for ``good'' and vanishes for ``bad'' sites (1):
I.e.\ for sites where the $k_{\rm F}$ conduction-electron wave function has a finite (vanishing) weight, the Kondo effect ``wins'' (RKKY wins).
The remaining unscreened moments are subjected to nonlocal RKKY exchange subsequently.
At a still lower energy scale $\delta$, determined by the residual coupling of the box to the environment, the Kondo scale becomes exponentially small again with a screening cloud leaking out into the environment (0) \cite{SA02}.
Here, for $J<J_{\delta}$, the RKKY coupling is dominant again \cite{GBA04,Sim05}. 
The arrow marks the range covered by our numerical calculations.
}
\label{fig:sc}
\end{figure}

Experimentally, Kondo boxes have been realized as an individual grain \cite{RBT95,RBT97}, as a single metallocene molecule \cite{BWD+05}, or as a small quantum dot which acts as a spin-half impurity and which is coupled to a large quantum dot with a finite level spacing \cite{KZC+06}. 
The $T_{\rm K}/\Delta$ ratio can be tuned by varying the voltage on the gates separating the two dots.
One-dimensional Kondo boxes can be realized by small Co clusters in short carbon nanotubes pieces \cite{OHCL00}
or by a carbon nanotube exchange coupled to a localized electron \cite{BBM+10}.

With recent progress in manipulation and characterization of magnetic systems on mesoscopic and nano scales, studies of the {\em competition} between Kondo screening and RKKY interaction in a quantum box come into reach as, e.g., in artificial and tunable double quantum-dot systems \cite{CTL+04}. 
Scanning-tunneling techniques nowadays allow to measure atomically precise maps of the RKKY coupling between individual adatom pairs on metallic surfaces \cite{ZWL+10} and the bottom-up construction of tailored magnetic nanostructures with atomic control \cite{KWCW11,KWC+12}.

\paragraph{Main results.}

Here we show that there is an unconventional competition between Kondo and RKKY that sets in for $J\ll J_{\rm D}$ in a system with several magnetic impurities coupled locally via $J$ to the sites of a finite quantum box (see Fig.\ \ref{fig:sc}).
This {\em re-entrant competition} is a consequence of a combination of two important finite-size effects becoming vital if $T_{K}^{\rm bulk} < \Delta$ or, equivalently, if $J< J_{\Delta}$:
First, there is an odd-even effect in the number of conduction electrons $N$. 
For even $N$, the Fermi energy lies in a finite-size gap (``off resonance''). 
As the Kondo scale is cut by $\Delta$, the local moments are unscreened but couple via the RKKY interaction. 
For odd $N$ (``on resonance''), however, cutting the Kondo scale implies that standard perturbation-theory in $J$ is regularized and that already the first-order-in-$J$ term leads to a singlet formation. 
This {\em linear}-in-$J$ Kondo scale wins over the RKKY scale $\propto J^{2}$ for weak $J$ if, secondly, the single-conduction-electron eigenstate at the Fermi edge has finite weight at the respective impurity site (``good site'') while for ``bad sites'' unscreened moments are subjected to RKKY coupling.
We demonstrate that a fairly complete qualitative picture of this physics is obtained by studying {\em three}-impurity Kondo models using exact diagonalization, perturbation theory and density-matrix renormalization group (DMRG)
\cite{Whi92,Sch10}.

For a completely isolated box the conduction-electron density of states consists of a set of delta-peaks separated by $\Delta$ in the vicinity of the Fermi edge.
A weak coupling of the box to an environment (e.g.\ leads), however, turns the delta-peaks into resonances with a characteristic width $\delta$ \cite{SA02}.
Thereby a fourth scale comes into play.
At this energy scale $\sim \delta \ll \Delta$, or equivalently for $J \sim J_{\delta} \ll J_{\Delta}$, the novel re-entrant competition in the Kondo-vs.-RKKY quantum box breaks down.
Our study thus bridges the gap between the conventional Kondo-vs.-RKKY physics for $J>J_{\Delta}$ and the regime $J<J_{\delta}$ studied previously \cite{GBA04,Sim05}.

\paragraph{Multi-impurity model.}
 
We consider $M$ spins $\ff S_{m}$, with spin-quantum numbers $1/2$, which are coupled locally via an antiferromagnetic exchange $J>0$ to the local spins $\ff s_{i}$ of a system of $N$ itinerant and non-interacting conduction electrons. 
An even total number $N+M$ of conduction electrons and localized spins $1/2$ is assumed such that 
a Fermi-liquid state with a total spin singlet can be reached for $L\to \infty$.

The conduction electrons hop with amplitude $t\equiv 1$ between non-degenerate orbitals and nearest-neighboring sites of a lattice with {\em finite} number of sites $L$.
The Hamiltonian is 
\begin{equation}
{\cal H} = - t \sum_{\langle i,j \rangle, \sigma} c^{\dagger}_{i\sigma} c_{j\sigma} + J \sum_{m=1}^{M} \ff s_{i_{m}} \ff S_{m} \: .
\label{eq:ham}
\end{equation}
Here, $c_{i\sigma}$ annihilates an electron at site $i=1,...,L$ with spin projection $\sigma=\uparrow, \downarrow$, and 
$\ff s_{i} = \frac{1}{2} \sum_{\sigma \sigma'} c^{\dagger}_{i\sigma} \ff \sigma_{\sigma\sigma'} c_{i\sigma'}$ is the local conduction-electron spin at $i$, where $\ff \sigma$ is the vector of Pauli matrices.
Impurity spins couple to the local conduction-electron spins at the sites $i_{m}$ where $m=1,...,M$.

For the sake of clarity, we consider a one-dimensional chain with open boundaries.
Diagonalization of the tight-binding part in Eq.\ (\ref{eq:ham}) yields nondegenerate conduction-band energies 
$\varepsilon_{k} = - 2t \cos k$ with discrete $k=\pi n / (L+1)$ labeled by integers $n=1,...,L$.
For a half-filled system, i.e.\ for $N=L$ electrons, this results in an energy-level spacing $\Delta = 2t \sin(\pi/(L+1))$ at the Fermi energy $\varepsilon_{\rm F}=0$ if $L$ is odd.
The local density of states at $i$ consists of a finite number of $\delta$-peaks only, $\rho_{ii}(\omega) = \sum_{k} U_{i k}^{2} \delta(\omega-(\varepsilon_{k} - \varepsilon_{\rm F}))$ where $U_{i k} = \sqrt{2/(L+1)} \sin (i \, k)$. 
Opposed to the continuum limit $L\to \infty$, it is no longer finite in the vicinity of $\varepsilon_{\rm F}$.

\paragraph{Effective RKKY model.}

For two impurity spins and even $N$, $\varepsilon_{\rm F}$ lies in a finite-size gap of the single-conduction-electron spectrum (``off-resonant case'').
There is a crossover, for finite $L$, from local Kondo-singlet formation at strong $J$ to nonlocal RKKY coupling of the spins for $J\to 0$ as can be seen in spin-correlation functions and susceptibilities \cite{LVK05,TSRP12}. 
Note that for $J\to 0$ (for $T_{\rm K}<\Delta$) the Kondo effect is absent, and free moments are generated:
The impurity spins cannot dynamically couple to the conduction-electron system as the Fermi sea is non-degenerate and thus a finite energy $\sim \Delta$ would be necessary to screen a spin.
Hence, the low-energy sector is exactly described by the effective RKKY two-spin model $H_{\rm RKKY} = - J_{12} \ff S_1 \ff S_2$ with $J_{\rm 12} \propto (-1)^{|i_{1}-i_{2}|} J^2 / |i_{1}-i_{2}|$ for $J\to 0$.
This standard RKKY physics is easily recovered for very small $L$ by full diagonalization of $\cal H$.

\paragraph{$M=3$ spins.}

The quantum box with three impurities is qualitatively different since a non-trivial competition between RKKY coupling and Kondo-singlet formation is possible {\em even for $J\to 0$}:
As $N$ is odd, the highest one-particle eigenenergy $\varepsilon_{k_{\rm F}}$ is singly occupied, and thus the ground state of the conduction-electron system is two-fold Kramers degenerate (``resonant case'').
Hence, screening of an impurity spin is possible for $J\to 0$ but competes with the RKKY exchange:
Consider a chain with one impurity spin coupled to the system via $J$ at the central site while the other two spins are coupled to the adjacent sites.
At second order in $J$, perturbation theory predicts indirect antiferromagnetic coupling of the central with the adjacent spins and ferromagnetic coupling between the latter.
A naive argument based on Doniach's idea would first consider the corresponding effective three-spin RKKY model which has a total impurity-spin doublet ground state (for any absolute magnitudes of the RKKY couplings).

However, it is in fact the Kondo effect that wins in this case:
Perturbation theory for the Kondo problem is regularized due to the finite-size gap $\Delta>0$ and predicts that, {\em if an impurity spin dynamically couples to the conduction-electron system} this happens on a linear-in-$J$ scale.
For sufficiently weak but finite $J$, this is larger than the RKKY scale $\propto J^{2}$ and we thus expect formation of a spin singlet involving the conduction-electron spins. 
We will call this a ``Kondo singlet'' although $T_{\rm K} \propto J$ rather than being exponentially small.

{\em Whether or not there is a perturbative coupling of the impurity spin}, depends on the weight factor $U_{ik}$ at $k=k_F$:
$U_{ik_F}$ is just the $i$-component of the conduction-band one-particle energy eigenstate at $\varepsilon_F$. 
At $k_F = \pi/2$, we have $U_{i k_F} = \sqrt{2/(L+1)} \sin (i \, k_F) \ne 0$ for $i=1,3,...,L-2,L$. 
We call these sites ``good'', while $U_{i k_F} = 0$ for ``bad'' sites $i=2,4,...,L-1$.
Assume that $L=4n+3$ with integer $n$, i.e.\ the central site is bad.
The physics of this system is thus dominated by a Kondo screening of one of the two outer impurities.
The RKKY interaction comes into play in a second step only and mediates an antiferromagnetic coupling between the two remaining spins which form a spin singlet. 
Hence, the ground state is a Kondo singlet entangled with an RKKY singlet,
$|{\rm GS} \rangle = 
| {\rm K}_{1} \rangle \otimes | {\rm RKKY}_{23}\rangle
-
| {\rm RKKY}_{12}\rangle \otimes  | {\rm K}_{3} \rangle$,
and the total spin $S_{\rm tot}=0$.

\paragraph{Low-energy model.}

This has in fact been verified by exact diagonalization for very small systems with $L=3$ and $L=7$ but also for somewhat larger chains with $L=19$ and $L=51$ using DMRG (see Ref.\ \cite{TSRP12} for methodical details).
Moreover, the ground-state symmetry and all spin-spin correlation functions are, up to order $J^{2}$, fully determined by an effective four-spin model (for $M=3$) that replaces the RKKY Hamiltonian
(see Supplemental Material \cite{suppl}):
\begin{equation}
H_{\rm eff} 
  = 
   \sum_{m=1}^M (J_m^{\rm (1)} + J_m^{\rm (2)}) \ff S_m \ff s_{F}
  - \sum_{m,n=1}^M J_{mn} \ff S_m \ff S_n \: .
\end{equation}
Here, the effective coupling constants depend on $U_{i_{m}k}$ and $\varepsilon_{k}$.
$\ff s_{F}$ is the spin of the fully delocalized $k_{F}$-electron -- the Kondo cloud extends over the entire system for $J\to 0$.
The ground-state symmetry crucially depends on the position of the impurities as both the linear, 
$J_m^{\rm (1)}
=
J |U_{i_mk_{\rm F}}|^2
$,
and the quadratic Kondo coupling,
$J_m^{\rm (2)}
=
2J^2 |U_{i_mk_{\rm F}}|^2 \sum_{p>k_{\rm F}} \frac{|U_{i_mp}|^2}{ \varepsilon_p - \varepsilon_{k_{\rm F}} }
$,
vanish at sites where $|U_{i_mk_{\rm F}}|^{2}=0$ while the RKKY coupling $J_{mn} \propto J^{2} \ne 0$.
 
\begin{figure}[t]
\centerline{\includegraphics[width=0.45\textwidth]{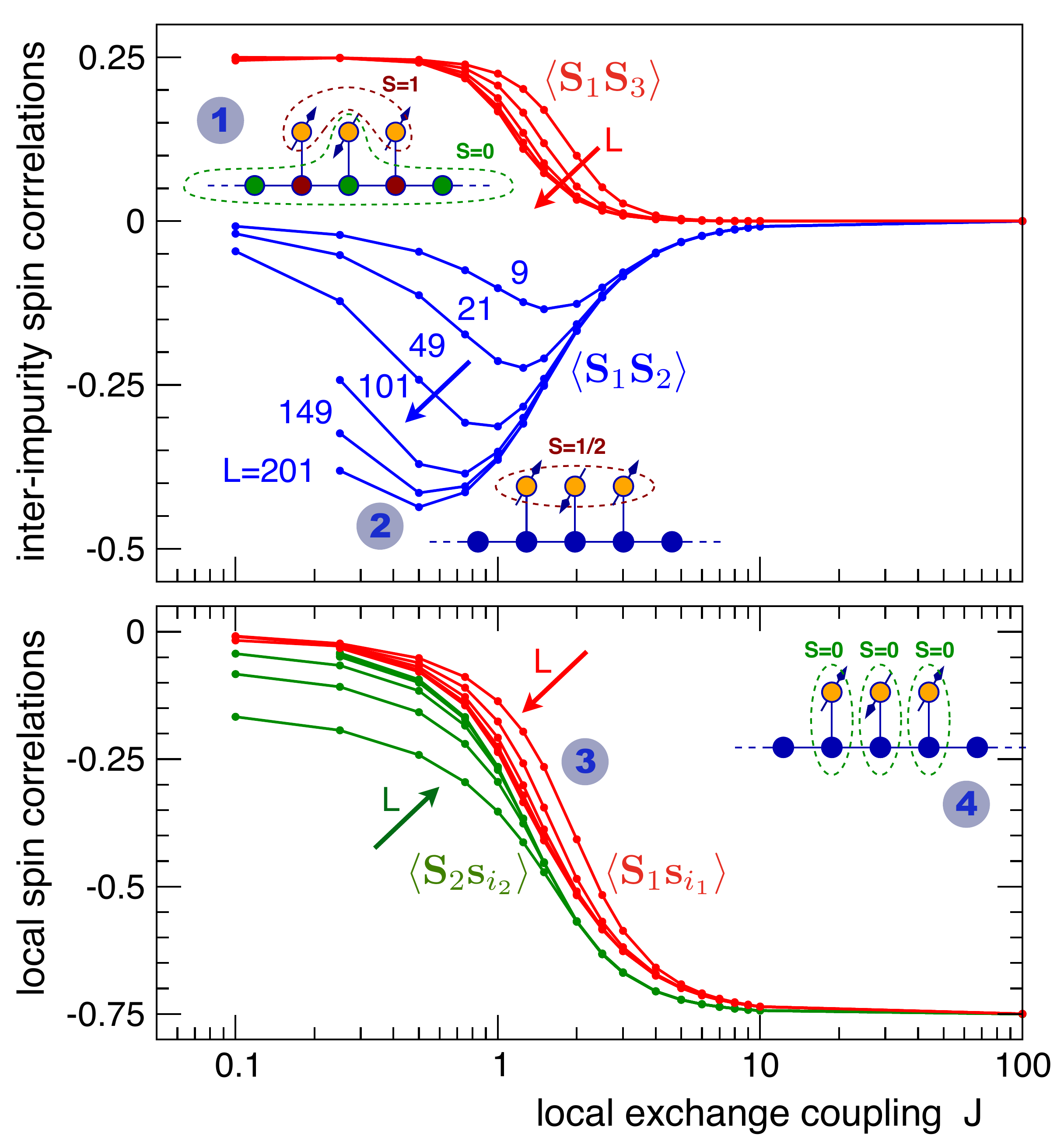}  }
\caption{(Color online) 
Inter-impurity and local spin correlations showing re-entrant competition between Kondo screening and RKKY coupling for the $M=3$ model (``bad-good-bad'') with increasing (see arrows) system size up to $L=201$ sites.
Note the log-scale for $J$.
With decreasing $J$, different regimes 1-4 (cf.\ Fig.\ \ref{fig:sc}) are found:
(4) Strong-coupling limit with local Kondo singlets.
(3) Kondo screening dominates over RKKY interaction.
(2) RKKY interaction is the leading energy scale for large systems ($J_{\rm D} > J_{\Delta}$).
(1) The linear Kondo scale dominates and leads to a screening of the central spin (coupled to a good site, green) while in a second step RKKY couples the remaining spins (at bad sites, red) to a nonlocal triplet.
}
\label{fig:spincorr}
\end{figure}

In a situation where the three local spins are all interacting with {\em bad} sites, for example, there is no Kondo effect. 
The inter-impurity distance is even which implies a ferromagnetic RKKY coupling leading to a total impurity spin $S_{\rm imp} = 3/2$ that does not couple to the total conduction-electron spin $S_{\rm cond}=1/2$ for weak $J$.
Hence, the total spin $S_{\rm tot}=1$ or $S_{\rm tot}=2$, the ground state is 8-fold degenerate.

If, at the same inter-impurity distance, all spins interact with {\em good} sites, the Kondo scale is large as compared to RKKY. 
Hence, one of the spins is Kondo-screened while the remaining ones are RKKY-coupled to a triplet: $S_{\rm tot} = 1$.
The ground state is an entangled Kondo-singlet-RKKY-triplet state.

\paragraph{Re-entrant competition.}

What happens for finite $\Delta$ as $J$ increases and what happens for finite $J$ if $\Delta$ decreases, i.e.\ if the system size $L$ increases?
Let us discuss those questions for the remaining case ``bad-good-bad'': 
Two spins $\ff S_{1}$ and $\ff S_{3}$ couple to bad sites neighboring the central site of a chain with $L=4n+1$. 
The central site is good. 
Therefore, for small $J$ and small $L$ we have $T^{\rm (bulk)}_{K} < \Delta$, the perturbative arguments given above apply, the Kondo scale is linear in $J$, and $\ff S_{2}$ is Kondo screened. 
The weaker (ferromagnetic) RKKY interaction then couples $\ff S_{1}$ and $\ff S_{3}$ to a nonlocal spin triplet, and thus $S_{\rm tot} =1$.

This is nicely reflected in the different ground-state spin-correlation functions obtained by DMRG which are shown in Fig.\ \ref{fig:spincorr}:
For $L=9$ at $J\to 0$, we find $\langle \ff S_{1} \ff S_{3} \rangle \to 1/4$ while $\langle \ff S_{1} \ff S_{2} \rangle \to 0$ (and $\langle \ff S_{2} \ff S_{3} \rangle \to 0$ by symmetry).
The local correlations vanish $\langle \ff S_{1} \ff s_{i_{1}} \rangle \to 0$ at the bad sites but for the good one $\langle \ff S_{2} \ff s_{i_{2}} \rangle$ remains finite as $J\to 0$.

The strong-$J$ limit is also easily understood:
For $J > J_{\Delta}$ the distinction between good and bad sites becomes irrelevant, and as $J\to \infty$ all local spin correlations tend to $-3/4$.
This indicates {\em local} Kondo-singlet formation which is basically independent of the system size.

At intermediate $J$, i.e.\ $J_{\Delta} < J < J_{\rm D}$, the conventional competition between Kondo screening and RKKY interaction should be recovered (since $J_{\Delta} < J$) , and the RKKY interaction should win (since $J<J_{\rm D}$).
However, this crucially depends on the system size: 
For small systems, we rather have $J_{\Delta} > J_{\rm D}$, and the intermediate-$J$ regime is skipped.
This can be seen in $\langle \ff S_{1} \ff S_{2} \rangle$ which stays close to zero in the entire $J$ regime with an only shallow minimum around $J=1$.

Only if the system is sufficiently large, namely if $J_{\Delta} < J_{\rm D}$ (i.e. $\Delta < J_{\rm RKKY}$), RKKY correlations among the impurity spins can develop (this is also the situation sketched in Fig.\ \ref{fig:sc}).
In fact, as is seen in Fig.\ \ref{fig:spincorr} for $L=201$, the correlation $\langle \ff S_{1} \ff S_{2} \rangle \to -1/2$ in the intermediate-$J$ regime and $\langle \ff S_{1} \ff S_{3} \rangle$ close to $1/4$. 
These are exactly the spin correlations of a three-spin system with ferromagnetic coupling between $\ff S_{1}$ and $\ff S_{3}$ and antiferromagnetic ones else.
With decreasing $J$, the system eventually crosses over to the perturbative regime at a $J_{\Delta}$ that strongly decreases with decreasing $L$.

One might speculate that the nonlocal impurity spin doublet that is formed by the RKKY interaction in the intermediate-$J$ regime,
$\Delta < T_{\rm K}^{\rm (bulk)} < J_{\rm RKKY}$, is Kondo screened in a subsequent step on a very small energy scale $T_{\rm K}^{\rm (3\, spins)} \ll T_{\rm K}^{\rm (bulk)}$. 
A conventional Kondo effect would be obtained only if $\Delta < T_{\rm K}^{\rm (3\, spins)} \ll T_{\rm K}^{\rm (bulk)}<J_{\rm RKKY}$, i.e.\ for very large systems.
Here, we rather expect $T_{\rm K}^{\rm (3\, spins)} \ll \Delta <  T_{\rm K}^{\rm (bulk)}<J_{\rm RKKY}$, i.e.\ the corresponding Kondo correlations are again cut by the system size and a linear-in-$J$ 3-spin Kondo scale develops. 
Still, for the maximum system size considered here ($L=201$), this scale appears too small to be resolved numerically.

Another issue for the quantum-box regime $T_{\rm K}^{\rm (bulk)} < \Delta$, is a weak coupling to the environment. 
Eventually, the total spin $S_{\rm tot} = 1$ of the ``bad-good-bad'' configuration, for example, may be screened by environmental spin degrees of freedom.
If the level broadening is small, $\delta \ll \Delta$, this is expected to take place on a very small energy scale $\ll \delta$ \cite{GBA04,Sim05}.
In the ``on resonance'' case, however, the RKKY interaction is particularly enhanced by intra-resonance processes \cite{Sim05}.

\paragraph{Conclusions.}

For an isolated quantum box including several localized magnetic moments coupled antiferromagnetically to conduction-electron spins, there is a subtle competition between Kondo screening and RKKY interaction even in the limit $J\to 0$.
The symmetry of the ground state crucially depends on the geometrical position of the impurities. 
It is predictable by perturbative techniques and experimentally accessible by studying field dependencies \cite{KWC+12}.
For sufficiently large systems ($J_{\rm D} > J_\Delta$), an {\em increasing} $J$ drives the system from Kondo screening on a linear-in-$J$ scale to conventional RKKY-coupled moments until the Kondo effect takes over again for large $J$.
The relevant parameter range can be estimated roughly by setting $T_{\rm K} \sim \Delta \sim J_{\rm RKKY}$:
RKKY couplings in the range of 0.1-10 meV correspond to Kondo temperatures of 1-100 K and, in a free-electron model \cite{TKvD99}, to system volumes of $(11.5 \mbox{nm})^3 - (2.5 \mbox{nm})^3$.

As one possible application, we suggest to employ quantum-confined systems at surfaces by means of scanning-tunneling techniques with the objective to construct nano-spintronics devices bottom-up \cite{KWCW11}. 
Confinement normal to the surface can be achieved e.g.\ by insulating spacers \cite{TMC06}, and lateral confinement will lead to strong variations in the local density of states as is known from quantum corrals, for example  \cite{CLEH95}, but also from nonmagnetic adatoms, step edges etc.
Even if the confinement is not perfect, the strong spatial dependence of the Kondo temperature is sufficient, if combined with an atomically precise positioning of atoms, to utilize the Kondo-vs.-RKKY physics in a quantum box. Different magnetic ground states for different spatial configurations of magnetic adatoms can be studied in real space as a function of an external magnetic field by means of single-atom magnetometry using spin-resolved scanning tunnelling spectroscopy \cite{KWC+12}.

The physics of the Kondo-vs.-RKKY quantum box is limited for strong $J$ by local Kondo-singlet formation and by residual couplings to the environment for very weak $J$. 
Generalizations to more impurities, systems off half-filling and multi-orbital systems are worth to be explored.
While one-dimensional models have been considered here for convenience, all main conclusions are expected to be valid in higher dimensions, too.
We also expect our work to serve as a bottom-up route to address dense or Kondo-lattice models.

\acknowledgments

We would like to thank I. Titvinidze for discussions.
Support of this work by the Deutsche Forschungsgemeinschaft (Sonderforschungsbereich 668, project A14) is gratefully acknowledged.

\end{document}